\documentclass[prd,amssymb,amsmath,amsfonts,floatfix,aps,nofootinbib,notitlepage,preprintnumbers]{revtex4-1}

\pdfoutput=1

\usepackage[T1]{fontenc}
\usepackage[utf8]{inputenc}

\usepackage{latexsym}
\usepackage{epsfig}
\usepackage{graphicx}
\usepackage{hyperref}
\hypersetup{pdfauthor={Kari Enqvist, Tomi Koivisto, Hannu J. Nyrhinen},pdftitle={Binary systems in Palatini-f(R) gravity},colorlinks=true,citecolor=blue,urlcolor=blue,linkcolor=blue}

\newcommand{\lp}{\left(}
\newcommand{\rp}{\right)}
\newcommand{\lb}{\left[}
\newcommand{\rb}{\right]}

\newcommand{\ba}{\begin{eqnarray}}
\newcommand{\ea}{\end{eqnarray}}
\newcommand{\be}{\begin{equation}}
\newcommand{\ee}{\end{equation}}

\newcommand{\bx}{{\bf x}}
\newcommand{\by}{{\bf y}}
\newcommand{\bv}{{\bf v}}

\newcommand{\al}{\alpha}

\newcommand{\R}{\hat{R}}
\newcommand{\Gammah}{\hat{\Gamma}}

\newcommand{\ud}{\: \mathrm{d}}

\newcommand{\M}{\mathcal{M}}
\newcommand{\I}{\overset{(3)}{\mathcal{I}^{kk}_{(eff)}}(t)}
\newcommand{\ti}{\tilde}
\newcommand{\f}{\varphi}
\usepackage{color}

\date{\today}

\begin{document}

\title{Binary systems in Palatini-$f(R)$ gravity}
\preprint{HIP-2013-10/TH}

\begin{abstract} 

We consider compact binary systems in $f(R)$ gravity theories in the Palatini approach and calculate the 
post-Newtonian parameters to the 1.5PN order
using the method of Direct Integration of the Relaxed Einstein equations (DIRE).  The Palatini-type modifications of gravity can be formulated as Einsteins gravity with 
modified response to matter sources, and it is shown in detail how to treat these correctly within the DIRE formalism. Our results explicitly confirm the expectation that  for binary black holes the new effects can be absorbed into redefinitions of the binary masses, rendering such systems
observationally identical to general relativity. 

\end{abstract}

\author{Kari Enqvist}
\email{kari.enqvist@helsinki.fi}
\author{Tomi Koivisto}
\email{tomik@astro.uio.no}
\author{Hannu J. Nyrhinen}
\email{hannu.nyrhinen@helsinki.fi}

\maketitle

\section{Introduction}

There is considerable interest in extensions of Einstein's general relativity (GR). Among the outstanding problems that GR alone is unable to address is the cosmological constant problem \cite{Weinberg:1988cp} at the infrared, and the singularities at the ultraviolet (for recent progress, see e.g. \cite{Biswas:2011ar,Modesto:2011kw}). Cosmological motivations for extensions include the observed acceleration of the expansion rate of the universe, which has inspired a plethora of gravity models that at best may be falsifiable only by their cosmological predictions \cite{Clifton:2011jh} for the data from probes such as the Euclid satellite \cite{Amendola:2012ys}. 

It is also useful to consider the possible implications of such models at other scales besides the cosmological, in particular in view of the classical solar system tests of gravity and experiments probing astrophysical phenomena. In future we might be able to detect gravitational wave signal from inspiralling compact binaries. The signal of the inspiral of supermassive black holes anywhere in the universe could in principle be detected by already now feasible space interferometer. In addition to the famous binary pulsar tests, black hole systems such as the quasar OJ287 at redshift $z=0.306$ \cite{Sillanpaa:1988zz} can provide crucial complementary constraints on extensions of GR. 

To study such systems, in practise one needs to resort to different approximation methods because of the extreme complexity of the relativistic field equations.  A canonical framework for this is the post-Newtonian (PN) formalism \cite{Will:2005va}, in which the metric and the equations of motion are  respectively expanded around the Minkowski metric and Newton's equations. Approximation in  PN orders  corresponds a series expansion in terms of  $(v/c)^2 \sim \epsilon \ll 1$, where $v$ is the velocity of the object and $c$ the speed of light. Here the first PN order corresponds to $(v/c)^2$, the second PN order corresponds to $ (v/c)^4$ etc. Testing gravity theories to higher accuracy requires the calculation of corrections to higher PN order. Future experiments may require accuracy even at the level $3.5$PN.

Brans-Dicke theories are prototype extensions of GR. Recently the full equations of motion for these theories were acquired up to 2.5PN order \cite{pati3}. As long as the scalar field can be considered massless, the results apply also for the $f(R)$ models of gravity \cite{sotiriou,Capozziello:2007ec}. It was confirmed that in scalar-tensor theories, for extended objects such as mixed black-hole neutron star systems, dipole radiation terms appear at 1.5PN order \cite{Damour:1995kt,Alsing:2011er}, but for binary black holes the form of the equations of motion stays unchanged. The difference between theories is that the masses of objects are rescaled by the scalar field. As there is no independent way to determine the masses, the system remains observationally indistinguishable from GR. However, as is well known, there are many ways around the restrictive assumptions underlying the no-hair theorems (for recent examples see e.g. \cite{Gregory:2013xca,Berti:2013gfa}), and therefore it is useful to explore the theory space for nontrivial observational signatures.

In this paper we study the PPN parameterisation of the Palatini-$f(R)$ theories. These theories constitute a very special class of scalar tensor theories in which the PPN parameters are identical to general relativity and the scalar field is nondynamical. Due to these peculiar properties and due to the presence of a potential term, the previous considerations of scalar field theories within the PPN context \cite{Damour:1995kt,Blanchet:1995fg,pati3} are not applicable to the Palatini theories. Nontrivial black hole solutions and other spherically symmetric spacetimes have been found in these theories  \cite{Kainulainen:2006wz,Barausse:2007pn,Olmo:2008pv,Olmo:2011np,Olmo:2012nx,Olmo:2012er}. Though their microscopic interpretation may be problematical \cite{Li:2008bma,Li:2008fa}, and while they may not produce viable alternatives to dark energy \cite{Koivisto:2005yc,Koivisto:2006ie} at least without warm dark matter \cite{Koivisto:2007sq}, it is of interest to understand their predictions for the binary systems. Indeed,  Palatini models may be seen as simplified versions of more realistic theories that unify the metric and the Palatini approaches \cite{Amendola:2010bk,Koivisto:2011vq,Harko:2011nh,Sandstad:2013oja} and can provide useful toy models for taking first steps to uncover the phenomenology of more complete theories. In such theories the PPN parameters can be different from GR already at the leading order \cite{Koivisto:2011tp}.

In the present paper we derive leading corrections to the equations of motion due to Palatini $f(R)$ theories using the DIRE approach of Pati, Will and Wiseman \cite{wiseman,pati1,pati3}. As a special case we consider a quadratic type of Palatini $f(R)$ theory and calculate the equations of motion of a compact binary system such as black holes at 1.5PN order. The paper is structured as follows: In section \ref{sec:direpalatini} we set up the necessary machinery for the DIRE approach in Palatini theories. In section \ref{sec:calculations} we calculate the components of the metric up to the order required for the equations of motion in a general $f(\R) = \R + \varphi(\R)$ theory. In section \ref{sec:eom} we substitute the results of the previous section to geodesic equations to obtain the PN equations of motion of a compact binary system. Section \ref{sec:conclusion} gives our conclusions. 

\section{\label{sec:direpalatini}DIRE approach for Palatini theories}

Let us first review the Palatini type of $f(R)$ theories of gravity and write their field equations in terms of an effective stress-energy tensor. This is convenient for implementation of the DIRE approach which is briefly sketched in section \ref{sec:dire}. The matter stress energy is specified in section \ref{matter}. Finally in section \ref{sec:metric} we compute the metric to the 1.5PN order. 

\subsection{Palatini $f(R)$ Theories of gravity}

The field equations of $f(R)$ type of gravity theories are given by the action \cite{sotiriou,Capozziello:2007ec}
\be
S = \frac{1}{2\kappa} \int \ud^4 x \sqrt{-g} f(R) + S_m,
\ee
where $\kappa= 8 \pi G$, $R = g^{\mu \nu} R_{\mu \nu}$ is the Ricci scalar contracted from the corresponding Ricci tensor with the inverse metric $g^{\mu \nu}$, $g$ is the determinant of the metric tensor and $S_m$ contains the matter parts of the action. $f(R)$ can in principle be any analytic function of the curvature scalar $R$. For $f(R) = R$ the action is the familiar Einstein-Hilbert one and the emerging theory is just GR.

In Palatini formulation of gravity the connection $\Gammah^\alpha _{\beta \gamma}$ and metric $g_{\mu \nu}$ are treated as independent variables. The field equations are found by varying with respect to both of these. In this case there are of course two sets of equations of motion describing the spacetime. 

In this approach the Ricci tensor is  defined as usual now only using the independent connection, i.e. $\R \equiv \R(\Gammah)$. From now on the hat denotes that a variable is defined using the independent connection $\Gammah^\alpha_{\beta \gamma}$. The hatted variables can depend also on the metric (e.g. $\R \equiv g^{\mu \nu} \R_{\mu \nu}$) while the hatless ones depend solely on the metric. The matter part of the Lagrangian, $S_m$ does not depend on the independent connection and matter follows the geodesics of the unhatted metric connection $\Gamma^\alpha_{\mu\nu}$ \cite{Koivisto:2005yk}. 

It is possible to find algebraic relations between the  metric curvature variables and the variables constructed of the independent connection. The relations can then in principle be used to write the field equations in terms of the metric curvature, which is convenient when we consider the equations of motion of compact objects. The field equations can be brought to Einstein-like form \cite{Koivisto:2005yc}
\ba \label{field_eq}
G_{\mu \nu} \equiv R_{\mu \nu} -  \frac{1}{2} g_{\mu \nu} R &=& \kappa T^{(eff)}_{\mu \nu},
\ea
where
\ba \kappa T^{(eff)}_{\mu \nu} &\equiv & \frac{\kappa}{f'(\R)} T_{\mu \nu} - \frac{g_{\mu \nu}}{2} \left( \frac{\kappa T + f(\R)}{f'(\R)}  \right) + \frac{1}{f'(\R)} \left(\nabla_\mu \nabla_\nu - g_{\mu \nu} \Box \right) f'(\R) - \nonumber \\
 && - \frac{3}{2} \frac{1}{f'(\R)^2} \lb \lp \nabla_\mu f'(\R) \rp \lp \nabla_\nu f'(\R) \rp - \frac{1}{2} g_{\mu \nu} \lp \nabla f'(\R)\rp^2 \rb. \label{eq:Teff}
\ea
As can be guessed, the effective stress-energy tensor $T^{(eff)}_{\mu \nu}$ will later take the place of the simple $T_{\mu \nu}$  in calculations. However, the curvature $\R$ is solved from the relation
\be \label{trace}
\R f'(\R)-2f(\R)=\kappa T\,,
\ee
that is simply the trace of the field equations (\ref{field_eq}) and where $T$ is trace of the usual matter stress tensor. 

Assuming an analytic action $f(\R)$ that admits a flat space solution, we can expand the gravitational corrections as
\be \label{quad} 
f(\R) = \R + \al \R^2 + \beta \R^3 + \dots \approx \R + \al \R^2\,.
\ee 
The higher order terms can be assumed to be important only at very high curvature regimes. In the following, when a specific form of  $f(\R)$ is required, we will take into account only the quadratic correction for simplicity.

\subsection{DIRE approach for Palatini $f(R)$ theories \label{sec:dire}}

Next we shall review the Direct Integration of Relaxed Einstein equations, or DIRE approach to compute the PPN expansion of the metric. This section section relies heavily on \cite{pati1} but most of the things on relaxed equations can also be found in \cite{gravitation}. The DIRE approach to calculating components of the metric uses a straightforward idea of solving the metric iteratively. The resulting components are then used to write the other relevant quantities such as connections. 

The Einstein equations $G_{\mu \nu} = \kappa T^{(eff)}_{\mu \nu}$ can be brought to relaxed form by first introducing a potential
\be \label{eq:h}
h^{\mu \nu} \equiv \eta^{\mu \nu} - \sqrt{-g} g^{\mu \nu}\,,
\ee
and choosing to use the harmonic gauge condition
\be
{h^{\mu \nu}}_{,\nu} = 0\,.
\ee
Now solving the metric tensor in \eqref{eq:h} and substituting into the Einstein equation gives the so called relaxed version of the field equations as equations for $h^{\mu \nu}$
\be \label{eq:relaxed}
\Box h^{\mu \nu} = -2 \kappa \tau^{\mu \nu},
\ee
where $\Box = -\partial/\partial t + \nabla^2$ is the flat-spacetime D'Alembertian and $\tau^{\mu \nu}$ is the effective stress-energy pseudotensor
\be \label{eq:tau}
\tau^{\mu \nu} = (-g)T^{\mu \nu}_{(eff)} + \frac{1}{16 \pi} \Lambda^{\mu \nu}\,.
\ee
Here $\Lambda^{\mu \nu}$ contains all the non-linear contribution of the potential $h^{\mu \nu}$ given by
\be \label{eq:lambda}
\Lambda^{\mu \nu} = 16\pi (-g) t^{\mu\nu}_{LL} + \lp {h^{\mu \alpha}}_{, \beta} {h^{\nu \beta}}_{,\alpha} - {h^{\mu \nu}}_{,\alpha \beta} h^{\alpha \beta} \rp\,
\ee
where $t^{\mu \nu}_{LL}$ is the \emph{Landau-Lifshitz pseudotensor}. 

It is possible to solve equation \eqref{eq:relaxed} formally as a functional of the source variables without specifying the motion of the source  (hence they are called ``relaxed'') \cite{pati1}. Note that formally the only difference with the Palatini case at hand and GR is that the stress energy tensor is replaced by the effective one as defined in Eq.(\ref{eq:Teff}).

In DIRE approach a formal solution of equation \eqref{eq:relaxed} is written as a retarded integral equation of $h^{\mu \nu}$. The integration runs over past null light cone, formally
\be \label{eq:integral}
h^{\mu \nu}(t,\bx) = 4 \int \frac{\tau^{\mu \nu}(t', \bx')\delta(t'-t+|\bx-\bx'|)}{|\bx-\bx'|}\ud^4 x'\,.
\ee
As can be seen from the definition of $\tau^{\mu \nu}$, both sides of the equation \eqref{eq:integral} contain $h^{\mu \nu}$. For this reason we solve the equation by iteration. We assume the slow-motion, weak-field approximation ($v < 1$, $||h^{\mu \nu}||< 1$) so that corrections get successively smaller on every iteration cycle.

The iteration of equation \eqref{eq:integral} proceeds as follows: first substitute $h^{\mu \nu} = 0$ to the right hand side of the equation \eqref{eq:integral} and solve for the first approximation $_1h^{\mu \nu}$. Next use $_1h^{\mu \nu}$  to get the effective stress-energy tensor $_1T^{\mu \nu}_{(eff)}$ (in terms of the ``conserved'' baryon density $\rho^*$, hence also $T^{\mu \nu}_{(eff)}$ gets corrections, see \eqref{eq:energytensor}) and $_1\Lambda^{\mu \nu}$ to this order.
Then use $_1T^{\mu \nu}_{(eff)}$ to solve the second-iterated $_2h^{\mu \nu}$. Continue until the required order is obtained. 

In the integrand of \eqref{eq:integral} $|\bx-\bx'|$  determines whether the observations are made in the so called near zone or radiation zone \cite{pati1}. The two zones are separated by the typical wave length $\mathcal{R}$ of gravitational radiation. For the purpose of studying equations of motion we assume that $|\bx-\bx'| \le 2 \mathcal{R}$. Points for which $|\bx-\bx'| \ge 2 \mathcal{R}$ are considered when gravitational waveforms are studied but for now we concentrate on the near zone.

\subsection{Matter} \label{matter}

Next we need to specify the matter sector. Throughout calculations we use the conserved baryon mass density $\rho^*$ instead of the locally measured one $\rho$. The relation between the two is $\rho^* \equiv \rho \sqrt{-g} u^0$.

We model matter by the pressureless perfect fluid stress-energy tensor
\be\label{eq:energytensor}
T^{\mu \nu} = \rho^* (-g)^{-1/2} u^0 v^\mu v^\nu\,,
\ee
where the velocity $v^i = u^i / u^0$. The definition of velocity is expanded to the time-like component so that $v^\mu = (1,v^i)$.

As Pati and Will \cite{pati2} we define the baryon rest mass, center of baryonic mass, velocity and acceleration of each body by
\begin{subequations} \label{eq:com}
\ba 
&m_A \equiv & \int_A \rho^* \ud^3x\,, \\
&\bx_A \equiv & \tfrac{1}{m_A} \int_A \rho^* \bx \ud^3 x\,, \\
&\bv_A \equiv & \frac{\ud \bx_A}{\ud t} = \tfrac{1}{m_A} \int_A \rho^* \bv \ud^3 x\,, \\
&\mathbf{a}_A \equiv & \frac{\ud \bv_A}{\ud t} = \tfrac{1}{m_A} \int_A \rho^* \mathbf{a} \ud^3 x \label{eq:a}\,.
\ea
\end{subequations}

We are then ready to proceed as sketched in the previous subsection.

\subsection{Metric tensor to 1.5 post-Newtonian order}
\label{sec:metric}

For the components of $h^{\alpha \beta}$ Pati \& Will \cite{pati1} define a simplified notation:
\begin{equation}
\begin{split}
N \equiv h^{00} \sim O(\epsilon)\,, \\
K^i \equiv h^{0i} \sim O(\epsilon^{3/2})\,,  \\
B^{ij} \equiv h^{ij} \sim O( f(\R) - \R) \,,  \\
B \equiv h^{ii} \equiv \sum_i h^{ii} \sim O(\epsilon) \,.
\end{split}
\end{equation}
We note that the order in which $B$ and $B^{ij}$ contribute to the equations does depend on the form of $f(\R)$, see \eqref{eq:hcomps}. In GR these terms are $O(\epsilon^2)$.  

The metric tensor can then be expressed in terms of the above components, formally $g_{\mu \nu} =  \sqrt{ -g}(\eta^{\alpha \beta} - h^{\alpha \beta})^{-1}$ (where -1 refers to matrix invert). To required order, which in calculating the equations of motion to 1.5PN order means $g_{00}$ to $O(\epsilon^{5/2})$, $g_{0i}$ to $O(\epsilon^{2})$ and $g_{ij}$ to $O(\epsilon^{3/2})$ 

\begin{subequations} \label{eq:metric}
\begin{align}
g_{00} = & - \left( 1 - \frac{1}{2} N + \frac{3}{8}N^2  \right)  + \frac{1}{2} B \lp 1 - \frac{1}{2} N - \frac{1}{4} B \rp + O(\epsilon^3)\,, \\[10pt]
g_{0i} = & -K^i + O(\epsilon^{5/2})\,,  \\[10pt]
g_{ij} = & \delta^{ij} \left(1 + \frac{1}{2} N - \frac{1}{2} B \right) + B^{ij} + O(\epsilon^2)\,. 
\end{align}
\end{subequations}
Also the determinant can be expanded, and to required order we have 
\be
(-g) = 1 + N + O(\epsilon^2).
\ee
Using these expansions it is a matter of substitution into \eqref{eq:lambda} to get the expression for the components of $\Lambda^{\alpha \beta}$. 
The result is given in \cite[eq. (4.4)]{pati1}.

In order to calculate the metric tensor we must find the components of $h^{\alpha \beta}$ to orders that in this case are $O(\epsilon^{5/2})$ for $N$ and $B$, $O(\epsilon^2)$ for $K^i$. Thus we use the notation 
\begin{subequations}\label{eq:components}
\begin{align}
N =& \epsilon \lp N_0 + \epsilon N_1 + \epsilon^{3/2} N_{1.5}   \rp + O(\epsilon^3)\,, \\
K^i =& \epsilon^{3/2} K^i_1  + O(\epsilon^{5/2})\,, \\
B =& \epsilon \lp B_0 + \epsilon B_1 + \epsilon^{3/2} B_{1.5} \rp + O(\epsilon^3)\,, \\
 B^{ij} =&  \epsilon^2 B^{ij}_1 + O(\epsilon^{5/2}).
\end{align}
\end{subequations}
For the components of $h^{\mu \nu}$, where the subscript on each term refers to post-Newtonian order of each term in the GR equations of motion. 

\section{\label{sec:calculations}Calculation of the metric components}

In this section we calculate the metric components following the procedure outlined above. First in \ref{n_order} we obtain the Newtonian results, and in \ref{pn_order} we compute the post-Newtonian corrections for $f(R)$ Palatini gravity.
\subsection{Newtonian order} 
\label{n_order}
As stated in \ref{sec:dire}, the iteration begins by assuming a flat space and setting $h^{\mu \nu} = 0$. Then the spacetime is Minkowskian, $R = 0 = \R$ (also $f(\R)=0$) and we assume that $f'(\R=0) = 1$. Then $\tau^{00} = (-g) T^{00} + O(\rho \epsilon) \approx \rho^* + O(\rho \epsilon)$. We get
\be \label{eq:N0}
N_0 ={} _0h^{00} = 4 \int_\M \frac{\rho^*}{|\bx - \bx'|} \ud^3 x' = 4 U,
\ee
where $U$ is the Newtonian potential and $\M$ denotes the intersection of a constant time hypersurface $t=\text{const}$ and the near-zone world tube $\mathcal{D} = \lbrace x^\mu : r < \mathcal{R}, - \infty < t < \infty  \rbrace$. The outer integrals vanish since we are dealing with compact objects inside the near zone.
This leads the metric determinant taking the form
\begin{equation}
(-g) = 1 + 4U + O(\epsilon^2)\,.
\end{equation}
To this order the metric is 
\begin{subequations}\label{eq:Newtng}
\begin{align}
g_{00}=& -1 + 2U +O(\epsilon^2)\,, \\
g_{0i}=& O(\epsilon^{3/2})\,, \\
g_{ij}=& \delta_{ij}(1 + 2U) + O(\epsilon^2)\,,
\end{align}
\end{subequations}
which is the usual result.

\subsection{Post-Newtonian order} 
\label{pn_order}

Let us consider a theory
\begin{equation}\label{fRphi}
f(\R) = \R + \f(\R)~,
\end{equation}
where $\f(\R)$ is some analytic function of the curvature scalar e.g. as in \eqref{quad}. For weak fields the theory should resemble GR so we also assume that $\f$ as well as its derivatives are small corrections compared to $\R$. Thus we can write the effective stress-energy tensor \eqref{eq:Teff} as
\be
T^{\mu \nu}_{(eff)} = \frac{T^{\mu \nu}}{f'} - g^{\mu \nu} \left( \frac{T}{2f'} + \frac{1}{16\pi} \frac{f}{f'}\right) + \frac{1}{8\pi f'} (\nabla^\mu \nabla^\nu - g^{\mu \nu} \Box)f'\,.  \\
\ee
That is (assuming as before that $\dot{f}(\R)=0$)
\begin{subequations}\ba
T^{00}_{(eff)}&=& \frac{\rho^*}{2f'} + \frac{f+2\nabla^2 f'}{16 \pi f'} +\epsilon \lb \frac{3}{4} \frac{\rho^* v^2}{f'}  + \frac{3\rho^* U}{2 f'}  - \frac{U(f+2\nabla^2 f')}{8 \pi f'} \rb + O(\epsilon^2)\,, \\ 
T^{0i}_{(eff)} &=& \frac{\rho^* v^i}{f'} + O(\epsilon^{3/2})\,, \\ 
T^{ij}_{(eff)} &=&  \frac{\nabla^i \nabla^j f'}{8 \pi f'} +\delta^{ij} \lp\frac{\rho^*}{2f'} - \frac{f+2\nabla^2 f'}{16 \pi f'} \rp  + \epsilon \lb \frac{\rho^* v^i v^j}{f'} - \delta^{ij} \lp \frac{\rho^* v^2 }{4f'} + \frac{\rho^* U}{2f'} + \frac{U(f+2\nabla^2 f')}{8 \pi f'} \rp \rb + O(\epsilon^2),
\ea \end{subequations}
where we have assumed $\dot{f}(\R) = 0$ and dropped all terms that are $O(\f^2)$.

The source term  \eqref{eq:tau} to the required order is then
\begin{subequations} \ba 
\tau^{00} &=& \frac{\rho^*}{2f'} + \frac{f+2\nabla^2 f'}{16 \pi f'} - \frac{7}{8 \pi} (\nabla U)^2 + \epsilon \lb \frac{3 \rho^* v^2}{4 f'} + \frac{7 \rho^* U}{2 f'} + \frac{U(f+2\nabla^2 f')}{8 \pi f'} \rb  + O(\rho^* \epsilon^2)\,, \\
\tau^{0i} &=& \frac{\rho^* v^i}{f'} + O(\rho^* \epsilon^{3/2})\,, \\
\tau^{ii} &=& \frac{\rho^*}{2f'} - \frac{f}{16 \pi f'}  -\frac{1}{8\pi} (\nabla U)^2 + \epsilon \lb \frac{3 \rho^* v^2}{4 f'} + \frac{3 \rho^* U }{2 f'} - \frac{U( 3 f + 2\nabla^2 f')}{8\pi f'} \rb + O(\rho^* \epsilon^2)\,, \\ 
\tau^{ij} &=& \frac{\nabla^i \nabla^j f'}{8 \pi f'} + O(\rho^* \epsilon)\,.
\ea \end{subequations}

Next we will calculate the near-zone integral \eqref{eq:integral}. For details of the evaluation, we refer the reader to  \cite{pati1}. 
The components of $h^{\alpha \beta}$ become (see Appendix A  for definitions of potentials) 
\begin{subequations} \label{eq:hcomps}\ba
N_0 + N_1 &=& 2 \ti U + \ti P(f)  + 7 U^2 + 3\ti \Phi_1 + 14 \ti \Phi_2 - 14 \Phi_2 +  2 \ti P(U f)  + 2 \ddot{\ti X}   -\frac{2\f'}{f'}  + 4 \ti P(U \nabla^2 \f')\,, \\
K_1 &=& 4\ti V^i\,, \\
B_0 + B_1 &=& 2 \ti U - \ti P(f) + U^2 + 3 \ti \Phi_1 + 6 \ti \Phi_2 - 2\Phi_2  - 6 \ti P(Uf) - 4 \ti P(U \nabla^2 \f')\,, \\
B^{ij} &=& 2 \ti P(\nabla^i \nabla^j \f')\,, \\
N_{1.5} &=& -\frac{2}{3}\I\,, \\
B^{ij}_{1.5} &=& -2 \I\,,
\ea\end{subequations}
where we use the fact that for Poisson potentials $P(\nabla^2 f) = -f +$boundary terms, given in Appendix D of \cite{pati1}. The last terms are the third time derivative of the $\mathcal{I}^Q$ variable given by 
\be
\mathcal{I^Q} \equiv \int_\M \tau^{00}_{(eff)} x^Q \ud^3 x,
\ee
where $Q\equiv i_1 \cdots i_k$ is given as for example $x^Q = x^{i_1 \cdots i_k} = x^{i_1} \cdots x^{i_k}$. The \~{} over a potential denotes scaling with $1/f'$ e.g.
\be 
\ti U \equiv \int_\M \frac{\rho^*/f'}{|\bx-\bx'|}\ud^3 x' \quad \text{and} \quad \ti \Phi_1 \equiv \int_\M \frac{\rho^* v^2/f'}{|\bx-\bx'|} \ud^3 x'.
\ee
The diagonal components $N$ and $B$ are divided into two parts according to order in which they contribute. Thus the subscript 0 denotes the parts
\ba  
N_0 &=& 2\ti U + \ti P(f) \nonumber \\
B_0 &=& 2 \ti U - \ti P(f)
\ea
and the rest is denoted by $N_1$ and $B_1$.

The components of $h^{\alpha\beta}$ can now be substituted to get $g_{\mu \nu}$, $\Gamma^\alpha_{\beta \gamma}$ and $a^i_{1PN}$.
First of all, the metric tensor to 1.5 PN order is
\begin{subequations} \ba \label{eq:g1.5}
g_{00} &=& -1 + 2\ti U + 4 U^2  - 3 \ti U^2  - \tfrac{1}{4}\ti P(f)^2 -  \ti U \ti P(f) + 3\ti \Phi_1 - 8 \Phi_2  + 10 \ti \Phi_2 - 2 \ti P(Uf)  + \ddot{\ti X}\,, \nonumber \\
&& - \frac{\f' ( 1 - 4\ti U - \ti P(f))}{f'} - \frac{4}{3}\I + O(\epsilon^3)\,,\\
g_{0i} &=&  4\ti V^i  + O(\epsilon^{5/2})\,,  \\
g_{ii} &=&  1 + 2\ti U - \frac{\f'}{f'}  + O(\epsilon^2)\,, \\
g_{ij} &=&  2 \ti P(\nabla^i \nabla^j \f') + O(\epsilon^2)\,. \label{eq:g1.5d}
\ea \end{subequations}

When $f(\R) = \R$ we have $f'=1$, $\f=0$, $P(f) = 2U$, $P(Uf) = 2\Phi_2$ and the above reduces to the 1.5PN metric of GR \cite{pati1}. When $f(\R)$ is something more complicated, we still should have $f'$ close to unity and $\f'$ and $P(\nabla^i \nabla^j \f')$ small enough to retain  \eqref{eq:Newtng} as the first approximation of $g_{\mu\nu}$.

\section{\label{sec:eom}Equations of Motion}

In this section we use the result of the previous section to obtain the equations of motion. First we review the geodesic equations of motion. In subsection \ref{conn} we deduce the connection and in subsection \ref{eoms} construct the equations of motion.

\subsection{Geodesic equations as equations of motion of matter}

In Palatini $f(R)$ gravity matter follows geodesics of the metric connection $\Gamma^\alpha_{\mu \nu} (g_{\mu \nu})$ (and not the independent connection $\Gammah^\alpha_{\mu \nu}$ as implied in some literature, e.g. \cite{Capozziello:2012eu}). The form of the used $f(R)$ theory affects the motion through the solution of equation \eqref{eq:Teff}, namely  $g_{\mu \nu}(\R)$  and hence through the metric connection $\Gamma(g_{\mu \nu})$. Thus, the equations of motion are given by the usual relation ${T^{\mu \nu}}_{;\nu} = 0$, which is equivalent to the geodesic equation $u^\mu {u^\nu}_{;\mu} = 0$.  With the coordinate time $t$ and the above definitions \eqref{eq:com} this can be brought into form  
\begin{equation} \label{eq:ai}
a^i = - \Gamma^i_{\alpha \beta} v^\alpha v^\beta + \Gamma^0_{\alpha \beta} v^\alpha v^\beta v^i\,,
\end{equation}
which is an equation for the coordinate acceleration of a fluid element. The total equations of motion of $A$th body are then given by integration over the body \eqref{eq:a}. For our case of two black holes we would like to be using the delta function density $\sim m_A \delta(\bx-\bx_A)$ so that the integral could be dealt with easily. However, for Palatini $f(R)$ problems may arise with terms that comprise the square of the baryon density $(\rho^*)^2$. Formally the result is given by 
\begin{equation}
a^i_A = \frac{1}{m_A} \int_A \rho^* \lp - \Gamma^i_{\alpha \beta} v^\alpha v^\beta + \Gamma^0_{\alpha \beta} v^\alpha v^\beta v^i \rp \ud^3 x\,,
\end{equation}
which is the equation of motion of the Ath body. Next we will expand the metric tensor and the metric connection to required order for our calculation.

\subsection{Metric connection expanded to required order}
\label{conn}

The metric connection is defined as 
\be
\Gamma^\alpha_{\mu \nu} \equiv  \frac{1}{2} g^{\alpha \lambda} \left( g_{\mu \lambda, \nu} + g_{\lambda \nu, \mu} - g_{\mu \nu, \lambda} \right).
\ee
An explicit expression in terms of components of $h^{\mu \nu}$ can be obtained by substituting the above expansions \eqref{eq:metric} and \eqref{eq:components} and using the result \eqref{eq:hcomps}. We get
\begin{subequations} \label{eq:connections}
\begin{align}
\Gamma^0_{00} =& - \epsilon \dot{\ti U} + O(\epsilon^2)\,, \\
\Gamma^0_{0i} =& - \epsilon \ti U^{,i} + O(\epsilon^2)\,, \\
\Gamma^0_{ij} =& O(\epsilon^2)\,, \\
\Gamma^i_{00} =& - \ti U^{,i} - \epsilon \lp \tfrac{1}{4} \lp N^{,i}_1 + B^{,i}_1 \rp + \dot{K}^i_1 -  5 \ti U \ti U^{,i} - \tfrac{1}{2} \lb \ti U \ti P(f)\rb^{,i} - \tfrac{1}{4} \ti P(f) \ti P^{,i}(f) \rp + O(\epsilon^2)\,, \\
\Gamma^i_{0j} =& \epsilon \lp \dot{\ti U} \delta^{ij} - K^{[i,j]}_1 \rp + O(\epsilon^2)\,, \\
\Gamma^i_{jk} =& \epsilon \, ^0\Gamma^i_{jk} (\ti U) + O(\epsilon^2)\,,
\end{align}
\end{subequations}
where 
\be
^0 \Gamma^{i}_{jk} (\ti U) \equiv \ti U^{,k} \delta^{ij} + \ti U^{,j} \delta^{ik} - \ti U^{,i} \delta^{jk}\,.
\ee
Apart from the scaling, the connection depends on the chosen theory only through $N_1$, $K_1$, $B_1$ and product terms. Thus any observable differences are at most in the post-Newtonian order as can be expected from the metric (\ref{eq:g1.5}-\ref{eq:g1.5d}). 

\subsection{Equations of motion}
\label{eoms}
 
To obtain the equations of motion we substitute the expanded connection \eqref{eq:connections} to \eqref{eq:ai}. The result is the coordinate acceleration in the form 
\be
\frac{\ud v^i}{\ud t} = a^i_N + a^i_{PN}\,.
\ee
We have \eqref{eq:ai} 
\ba
a^i &=& \ti U^{,i} + \epsilon \lp 4UU^{,i} - 5 \ti U \ti U^{,i} - \tfrac{1}{2} \lb \ti U \ti P(f) \rb^{,i} -\tfrac{1}{4} \ti P(f) \ti P^{,i}(f) + \tfrac{3}{2} \ti \Phi_1^{,i} - 4 \Phi_2^{,i} + 5 \ti \Phi_2^{,i} - \ti P^{,i}(Uf)  + \right.\nonumber \\
&&\left. +  \tfrac{1}{2} \ddot{\ti X}^{,i} + 4 \dot{\ti V}^i. - 3 \dot{\ti U} v^i + 8 \ti V^{[i,j]} v^j - 4 \ti U^{,j} v^j v^i + \ti U^{,i} v^2 - \lp \frac{\f'}{2f'} \rp^{,i}  \rp\,,
\ea
which gives the Newtonian acceleration of the 1st body as
\ba
(a^i_1)_N = \frac{-1}{m_1} \int_1 \ud^3 x \rho^* \int_\M \ud x' \frac{{\rho^*}'}{f'} \frac{x^i - {x'}^i}{|\bx - \bx'|^3} = -\frac{\ti m_2 n^i}{r^2}\,. 
\ea

In the next iteration cycle the apparent differences in the potentials vanish. The terms originating from $\Lambda^{\mu\nu}$ get the $1/f'$ scaling and $B$ dependency. Also the product terms with $\ti P(f)$ vanish due to modifications in $\Lambda^{\mu \nu}$ and we are left with $-4\ti U \ti U^{,i}$.

\paragraph{$f(\R) = \R + \al \R^2$} We are next going to concentrate to a theory of the form $f(\R) = \R + \alpha \R^2$, which we interpret as being an approximation of a more general theory as in (\ref{quad}). 
We also consider $\alpha \R$ as a small expansion parameter and expand the scaling factor $1/f'$ around $\alpha \R = 0$ as
\be
(f')^{-1} = 1 - 2\alpha\R.
\ee
The trace equation (\ref{trace})  retains its GR form
\be \label{eq:R(T)solved}
\R = -8\pi T\,.
\ee
Then using an appropriate approximation for  $T(\rho^*)$ we have 
\be \R = - 8 \pi \rho^* \lp - 1 + \epsilon (\tfrac{1}{2} v^2 + 3 U) \rp.\ee

We then need to specify the density distribution.

For most cases we might just as well use the delta distribution density $m_A \delta(\bx_A - \bx)$. However for the Palatini correction term $ \sim \int \rho^2 \ud x$ this is not mathematically well defined. Whatever form of distribution is used we always seem to end up in problems in the leading term in the limit where $\rho^*\to \infty$.  It is a well known property of Palatini gravity that the Newtonian limit depends on the density of the sources \cite{Olmo:2005zr,sotiriou}. Using a delta-distribution for density can result in unphysical pathologies whereas such a distribution could work as a good approximation in GR. We therefore take into account a finite size of the compact bodies, and call the remaining density the characteristic density of the $i$th body, corresponding to characteristic size $s$, ie. $\rho^*_i \sim m_i/s_i^3$. For black holes we can then consider the limit where $s$ vanishes.

The leading Palatini terms, the ones that are $O(\epsilon^{-3})$ can be obtained using similar kind of logic as the usual general relativistic ones. For these we get for example
\ba
a^i_1 (U_P) &=& 2\alpha \frac{m_2 \rho^*_2 n^i}{r^2}\,, \\
a^i_1 (V^{[i,j]}_P) &=& 2\alpha \frac{m_2 \rho^*_2}{2 r^2} \lp v^i_2 (\bv_1 \cdot n) -n^i (\bv_1 \cdot \bv_2) \rp\,, 
\ea
where $\rho^*_2 \sim m_2/s^3$ is the characteristic density of the second body. 

These Palatini-terms have the exact same form as the ones coming from $U^{,i}$ and $v^j V^{[i,j]}$. The only difference is that the Palatini corrections are multiplied by $2\alpha (8 \pi \rho^*)$, which is  not a surprising form after we evaluated $(1+2\alpha R)^{-1} \approx 1 - 2\alpha \R$. Thus the the equations of motion of a body retain their form but one has to take into account a scaling of mass of the other body.  This is in line with \cite{pati3} in which the authors find that for scalar-tensor theories the equations of motion are indistinguishable to the general-relativistic ones. Their form does not change and the only difference is in the coupling constant, which scales with the scalar field. However, one could doubt whether our approximation of zero radius is justified. Next we look at this point more carefully, since the treatment of sources is crucial in the Palatini-type theories.

\subsubsection{Expanding around $s=0$}

The logic goes as follows: We make a coordinate transformation from global harmonic coordinates $x^i$ to spatial coordinates $\hat{x}^i$ so that $\hat{\bx}_A$ is the position measured from the baryonic center of mass of the Ath body.
\be
x^i = x^i_A + \hat{x}^j \delta^i_j + O(\epsilon),
\ee 
where the correction terms of order $O(\epsilon)$ comprise the effects such as Lorentz boost. We then expand the integrands in powers of the characteristic size of the bodies $|\hat{x}| \sim s$. We use the general formula 
\be \label{eq:sseries}
\frac{1}{|\bx + \by|} = \sum_{q=0} ^\infty \frac{y^Q}{q!} \nabla^Q \lp \frac{1}{r} \rp, \qquad \text{where } |y|<|x| \equiv r. 
\ee
and finally we drop all remaining terms that depend on $s$ that is, are $O(s^n)$ for any $n\in \mathbb{Z}\setminus 0$ ($n$ can be negative). In other words we only keep terms that are $O(s^0)$. 

Also in our case the density can be thought to be distributed in the characteristic volume of each body, that is $\rho^* \sim m_A / s^3$. Thus the leading term in the \eqref{eq:sseries} series is actually of order $O(s^{-3})$. By \cite{pati2} the negative powers of $s$ correspond to self-energy corrections of the body and are omitted in the original paper. However, it is known that the gravitational constant depends on the scalar field which leads the observed mass to also depend on the scalar field. By analogy between Palatini $f(R)$ theories and the Brans-Dicke (or more general scalar-tensor) theory a correction like this could have been expected.

To get rid of $s$ we have to use term of appropriate power in  the series \eqref{eq:sseries}, namely the term with three $s^i$ vectors. The $s^i$ is vector that similar to $(x-x')$. Its  size is of order $s$ and it lies inside the body.  All terms with odd number of contributions of $s^i$ vanish because of symmetry \cite[p.10]{pati2}. For example,  by expanding the denominator as \eqref{eq:sseries} we get for PN acceleration due to the Palatini-term 
\ba
a^i_1 (\Phi^{,i}_P) &=& \frac{-10}{m_1} \int_1 \ud^3 x \rho^* \int_\M \ud^3 x' (\rho'^*)^2 \frac{(x-x')^i \lp \hat{x}^j(x-x')^j \rp^3}{|\bx-\bx'|^{9}} \\
a^i_1 (V^{[i,j]}_P) &=& \frac{-10}{m_1} \int_1 \ud^3 x \rho^* \int_\M \ud^3 x' (\rho'^*)^2 \frac{v'^{[i}(x-x')^{j]} \lp \hat{x}^k(x-x')^k \rp^3}{|\bx-\bx'|^{9}} \\
\ea
which seem to vanish due to spherical symmetry. Thus we are left with only the self-energy correction.

\subsubsection{The acceleration equation}

All in all the acceleration to 1PN order with the leading Palatini corrections is 
\ba 
a^i_1 & = & \frac{\ti m_2}{r^2}n^i \lp -1 + 4 \frac{\ti m_2}{r} + 5\frac{\ti m_1}{r} -v_1^2 + 4 v_1 \cdot v_2  - 2 v_2^2 + \frac{3}{2}	(v_2 \cdot n)^2 \rp  + \frac{\ti m_2}{r^2} (v_1-v_2)^i (4v_2\cdot n - 3 v_1 \cdot n).
\ea
This completes our computation.

\section{\label{sec:conclusion}Conclusions}

We have computed the acceleration equations for compact binary systems in Palatini-$f(R)$ theories up to leading order in the Post-Newtonian expansion using the DIRE method. Since these theories consist of a singular class of scalar-tensor theories in which the scalar field is non-dynamical, this exercise contributes a missing detail to the general understanding of the PPN behaviour of scalar-tensor theories. It also clarifies important aspects of Palatini theories which have been subjects of long-standing debates in the literature and concern the equivalence of the conformal frames \cite{Flanagan:2003rb,Vollick:2003ic},  the equivalence principle \cite{Koivisto:2005yk,Olmo:2006zu}, the interpretation of averaged equations \cite{Li:2008bma,Li:2008fa} and the viability of polytrophic sphere solutions \cite{Barausse:2007pn,Olmo:2008pv}. 

We found that the equations of motion retain their form in the presence of Palatini-$f(R)$ type corrections to GR. Only the masses of the objects have to be rescaled since the corrections can be re-expressed as a modified response of gravity to matter sources. Since there is no way to independently determine the masses of the known binary objects, the Palatini models are observationally indistinguishable from GR and one has to use e.g. cosmological data to constrain them. Our result may have been expected since it is known that in vacuum these Palatini theories reduce to Einstein's GR with a possible cosmological constant (depending upon the form of the function $f(R)$). Nevertheless, new black hole solutions with hair have been presented in the literature for these theories (e.g. \cite{Olmo:2012er}), and their equations of motion for the binary system could not have been deduced without the explicit and rather nontrivial calculation presented here.  

It remains to be seen how generically the scalar tensor theories reduce to (rescaled) GR predictions in the PPN limit. Thus far studies have focused upon the Brans-Dicke form of couplings that are conformally equivalent to GR. The most general viable scalar-tensor theory, the so called Horndeski action, however includes other coupling terms which are related to GR via disformal transformations. It would be interesting to see whether these coupling terms could be constrained by their possible effects to the dynamics of compact binary objects.

\acknowledgements
TK is supported by the Research Council of Norway, HN is supported by Magnus Ehrnrooth foundation, and
KE by the Academy of Finland grants 1218322 and  1263714;

\appendix
\section{List of Newtonian and  PN potentials used}
Here we list the definitions of potentials given in Appendix A of \cite{pati2} that we used.

All of them are some Poisson-like potentials or closely related to those.
\ba
P(f) & \equiv & \frac{1}{4\pi} \int_\M \frac{f(t,\bx')}{|\bx-\bx'|} \ud^3 x' \nonumber \\
S(f) & \equiv & \frac{1}{4\pi} \int_\M {f(t,\bx')}{|\bx-\bx'|} \ud^3 x' 
\ea

All of the potentials used are defined in terms of the conserved baryon density $\rho^*$
\ba
\Sigma(f) &\equiv & \int_\M \frac{\rho^*(t,\bx') f(t,\bx,)}{|\bx-\bx'|} \ud^3 x' = P(4\pi \rho^* f) \nonumber \\
X(f) &\equiv & \int_\M \rho^*(t,\bx') f(t,\bx') |\bx-\bx'| \ud^3 x' = S(4\pi \rho^*f)
\ea

Especially we use the PN potentials
\ba 
& U \equiv \Sigma(1), \quad V^i \equiv \Sigma(v^i) \quad \Phi^{ij}_1 \equiv \Sigma(v^i v^j), \quad  \Phi_1 \equiv \Sigma(v^2), \quad \Phi_2 \equiv \Sigma(U), \quad \nonumber\\
& X \equiv X(1), \quad U_P \equiv \Sigma(\rho^*), \quad V^i_P \equiv \Sigma(\rho^* v^i),
\ea
where the subscript $P$ denotes a term that does not appear in the GR equations.

\bibliography{references.bib}

\begin{thebibliography}{10}%
\makeatletter
\providecommand \@ifxundefined [1]{%
 \ifx #1\undefined \expandafter \@firstoftwo
 \else \expandafter \@secondoftwo
\fi
}%
\providecommand \@ifnum [1]{%
 \ifnum #1\expandafter \@firstoftwo
 \else \expandafter \@secondoftwo
\fi
}%
\providecommand \enquote [1]{``#1''}%
\providecommand \bibnamefont  [1]{#1}%
\providecommand \bibfnamefont [1]{#1}%
\providecommand \citenamefont [1]{#1}%
\providecommand\href[0]{\@sanitize\@href}%
\providecommand\@href[1]{\endgroup\@@startlink{#1}\endgroup\@@href}%
\providecommand\@@href[1]{#1\@@endlink}%
\providecommand \@sanitize [0]{\begingroup\catcode`\&12\catcode`\#12\relax}%
\@ifxundefined \pdfoutput {\@firstoftwo}{%
 \@ifnum{\z@=\pdfoutput}{\@firstoftwo}{\@secondoftwo}%
}{%
 \providecommand\@@startlink[1]{\leavevmode\special{html:<a href="#1">}}%
 \providecommand\@@endlink[0]{\special{html:</a>}}%
}{%
 \providecommand\@@startlink[1]{%
  \leavevmode
  \pdfstartlink
   attr{/Border[0 0 1 ]/H/I/C[0 1 1]}%
   user{/Subtype/Link/A<</Type/Action/S/URI/URI(#1)>>}%
  \relax
 }%
 \providecommand\@@endlink[0]{\pdfendlink}%
}%
\providecommand \url  [0]{\begingroup\@sanitize \@url }%
\providecommand \@url [1]{\endgroup\@href {#1}{\urlprefix}}%
\providecommand \urlprefix [0]{URL }%
\providecommand \Eprint[0]{\href }%
\@ifxundefined \urlstyle {%
  \providecommand \doi [1]{doi:\discretionary{}{}{}#1}%
}{%
  \providecommand \doi [0]{doi:\discretionary{}{}{}\begingroup
  \urlstyle{rm}\Url }%
}%
\providecommand \doibase [0]{http://dx.doi.org/}%
\providecommand \Doi[1]{\href{\doibase#1}}%
\providecommand \bibAnnote [3]{%
  \BibitemShut{#1}%
  \begin{quotation}\noindent
    \textsc{Key:}\ #2\\\textsc{Annotation:}\ #3%
  \end{quotation}%
}%
\providecommand \bibAnnoteFile [2]{%
  \IfFileExists{#2}{\bibAnnote {#1} {#2} {\input{#2}}}{}%
}%
\providecommand \typeout [0]{\immediate \write \m@ne }%
\providecommand \selectlanguage [0]{\@gobble}%
\providecommand \bibinfo [0]{\@secondoftwo}%
\providecommand \bibfield [0]{\@secondoftwo}%
\providecommand \translation [1]{[#1]}%
\providecommand \BibitemOpen[0]{}%
\providecommand \bibitemStop [0]{}%
\providecommand \bibitemNoStop [0]{.\EOS\space}%
\providecommand \EOS [0]{\spacefactor3000\relax}%
\providecommand \BibitemShut [1]{\csname bibitem#1\endcsname}%
\bibitem{Weinberg:1988cp}%
  \BibitemOpen
  \bibfield{author}{%
  \bibinfo {author} {\bibfnamefont{S.}~\bibnamefont{Weinberg}},\ }%
  \bibfield{journal}{%
  \Doi{10.1103/RevModPhys.61.1}{\bibinfo {journal} {Rev.Mod.Phys.}}\ }%
  \textbf{\bibinfo {volume} {61}},\ \bibinfo {pages} {1} (\bibinfo {year}
  {1989})%
  \bibAnnoteFile{NoStop}{Weinberg:1988cp}%
\bibitem{Biswas:2011ar}%
  \BibitemOpen
  \bibfield{author}{%
  \bibinfo {author} {\bibfnamefont{T.}~\bibnamefont{Biswas}}, \bibinfo {author}
  {\bibfnamefont{E.}~\bibnamefont{Gerwick}}, \bibinfo {author}
  {\bibfnamefont{T.}~\bibnamefont{Koivisto}},\ and\ \bibinfo {author}
  {\bibfnamefont{A.}~\bibnamefont{Mazumdar}},\ }%
  \bibfield{journal}{%
  \Doi{10.1103/PhysRevLett.108.031101}{\bibinfo {journal} {Phys.Rev.Lett.}}\ }%
  \textbf{\bibinfo {volume} {108}},\ \bibinfo {pages} {031101} (\bibinfo {year}
  {2012}),\ \Eprint{http://arxiv.org/abs/1110.5249}{arXiv:1110.5249 [gr-qc]}%
  \bibAnnoteFile{NoStop}{Biswas:2011ar}%
\bibitem{Modesto:2011kw}%
  \BibitemOpen
  \bibfield{author}{%
  \bibinfo {author} {\bibfnamefont{L.}~\bibnamefont{Modesto}},\ }%
  \bibfield{journal}{%
  \Doi{10.1103/PhysRevD.86.044005}{\bibinfo {journal} {Phys.Rev.}}\ }%
  \textbf{\bibinfo {volume} {D86}},\ \bibinfo {pages} {044005} (\bibinfo {year}
  {2012}),\ \Eprint{http://arxiv.org/abs/1107.2403}{arXiv:1107.2403 [hep-th]}%
  \bibAnnoteFile{NoStop}{Modesto:2011kw}%
\bibitem{Clifton:2011jh}%
  \BibitemOpen
  \bibfield{author}{%
  \bibinfo {author} {\bibfnamefont{T.}~\bibnamefont{Clifton}}, \bibinfo
  {author} {\bibfnamefont{P.~G.}\ \bibnamefont{Ferreira}}, \bibinfo {author}
  {\bibfnamefont{A.}~\bibnamefont{Padilla}},\ and\ \bibinfo {author}
  {\bibfnamefont{C.}~\bibnamefont{Skordis}},\ }%
  \bibfield{journal}{%
  \Doi{10.1016/j.physrep.2012.01.001}{\bibinfo {journal} {Phys.Rept.}}\ }%
  \textbf{\bibinfo {volume} {513}},\ \bibinfo {pages} {1} (\bibinfo {year}
  {2012}),\ \Eprint{http://arxiv.org/abs/1106.2476}{arXiv:1106.2476
  [astro-ph.CO]}%
  \bibAnnoteFile{NoStop}{Clifton:2011jh}%
\bibitem{Amendola:2012ys}%
  \BibitemOpen
  \bibfield{author}{%
  \bibinfo {author} {\bibfnamefont{L.}~\bibnamefont{Amendola}} \emph{et~al.}
  (\bibinfo {collaboration} {Euclid Theory Working Group})}%
   (\bibinfo {year} {2012}),\
  \Eprint{http://arxiv.org/abs/1206.1225}{arXiv:1206.1225 [astro-ph.CO]}%
  \bibAnnoteFile{NoStop}{Amendola:2012ys}%
\bibitem{Sillanpaa:1988zz}%
  \BibitemOpen
  \bibfield{author}{%
  \bibinfo {author} {\bibfnamefont{A.}~\bibnamefont{Sillanpaa}}, \bibinfo
  {author} {\bibfnamefont{S.}~\bibnamefont{Haarala}}, \bibinfo {author}
  {\bibfnamefont{M.}~\bibnamefont{Valtonen}}, \bibinfo {author}
  {\bibfnamefont{B.}~\bibnamefont{Sundelius}},\ and\ \bibinfo {author}
  {\bibfnamefont{G.}~\bibnamefont{Byrd}},\ }%
  \bibfield{journal}{%
  \bibinfo {journal} {Astrophys.J.}\ }%
  \textbf{\bibinfo {volume} {325}},\ \bibinfo {pages} {628} (\bibinfo {year}
  {1988})%
  \bibAnnoteFile{NoStop}{Sillanpaa:1988zz}%
\bibitem{Will:2005va}%
  \BibitemOpen
  \bibfield{author}{%
  \bibinfo {author} {\bibfnamefont{C.~M.}\ \bibnamefont{Will}},\ }%
  \bibfield{journal}{%
  \bibinfo {journal} {Living Rev.Rel.}\ }%
  \textbf{\bibinfo {volume} {9}},\ \bibinfo {pages} {3} (\bibinfo {year}
  {2006}),\ \Eprint{http://arxiv.org/abs/gr-qc/0510072}{arXiv:gr-qc/0510072
  [gr-qc]}%
  \bibAnnoteFile{NoStop}{Will:2005va}%
\bibitem{pati3}%
  \BibitemOpen
  \bibfield{author}{%
  \bibinfo {author} {\bibfnamefont{S.}~\bibnamefont{Mirshekari}}\ and\ \bibinfo
  {author} {\bibfnamefont{C.~M.}\ \bibnamefont{Will}},\ }%
  \bibfield{journal}{%
  \bibinfo {journal} {Phys. Rev. D},\ \bibinfo {pages} {submitted}}%
   (\bibinfo {year} {2013}),\
  \Eprint{http://arxiv.org/abs/arXiv/1301.4680}{arXiv/1301.4680}%
  \bibAnnoteFile{NoStop}{pati3}%
\bibitem{sotiriou}%
  \BibitemOpen
  \bibfield{author}{%
  \bibinfo {author} {\bibfnamefont{T.~P.}\ \bibnamefont{Sotiriou}}\ and\
  \bibinfo {author} {\bibfnamefont{V.}~\bibnamefont{Faraoni}},\ }%
  \bibfield{journal}{%
  \bibinfo {journal} {Rev. Mod. Phys.}\ }%
  \textbf{\bibinfo {volume} {82}},\ \bibinfo {pages} {451} (\bibinfo {year}
  {2010}),\ \Eprint{http://arxiv.org/abs/0805.1726v4}{arXiv:0805.1726v4
  [gr-qc]}%
  \bibAnnoteFile{NoStop}{sotiriou}%
\bibitem{Capozziello:2007ec}%
  \BibitemOpen
  \bibfield{author}{%
  \bibinfo {author} {\bibfnamefont{S.}~\bibnamefont{Capozziello}}\ and\
  \bibinfo {author} {\bibfnamefont{M.}~\bibnamefont{Francaviglia}},\ }%
  \bibfield{journal}{%
  \Doi{10.1007/s10714-007-0551-y}{\bibinfo {journal} {Gen.Rel.Grav.}}\ }%
  \textbf{\bibinfo {volume} {40}},\ \bibinfo {pages} {357} (\bibinfo {year}
  {2008}),\ \Eprint{http://arxiv.org/abs/0706.1146}{arXiv:0706.1146
  [astro-ph]}%
  \bibAnnoteFile{NoStop}{Capozziello:2007ec}%
\bibitem{Damour:1995kt}%
  \BibitemOpen
  \bibfield{author}{%
  \bibinfo {author} {\bibfnamefont{T.}~\bibnamefont{Damour}}\ and\ \bibinfo
  {author} {\bibfnamefont{G.}~\bibnamefont{Esposito-Farese}},\ }%
  \bibfield{journal}{%
  \Doi{10.1103/PhysRevD.53.5541}{\bibinfo {journal} {Phys.Rev.}}\ }%
  \textbf{\bibinfo {volume} {D53}},\ \bibinfo {pages} {5541} (\bibinfo {year}
  {1996}),\ \Eprint{http://arxiv.org/abs/gr-qc/9506063}{arXiv:gr-qc/9506063
  [gr-qc]}%
  \bibAnnoteFile{NoStop}{Damour:1995kt}%
\bibitem{Alsing:2011er}%
  \BibitemOpen
  \bibfield{author}{%
  \bibinfo {author} {\bibfnamefont{J.}~\bibnamefont{Alsing}}, \bibinfo {author}
  {\bibfnamefont{E.}~\bibnamefont{Berti}}, \bibinfo {author}
  {\bibfnamefont{C.~M.}\ \bibnamefont{Will}},\ and\ \bibinfo {author}
  {\bibfnamefont{H.}~\bibnamefont{Zaglauer}},\ }%
  \bibfield{journal}{%
  \Doi{10.1103/PhysRevD.85.064041}{\bibinfo {journal} {Phys.Rev.}}\ }%
  \textbf{\bibinfo {volume} {D85}},\ \bibinfo {pages} {064041} (\bibinfo {year}
  {2012}),\ \Eprint{http://arxiv.org/abs/1112.4903}{arXiv:1112.4903 [gr-qc]}%
  \bibAnnoteFile{NoStop}{Alsing:2011er}%
\bibitem{Gregory:2013xca}%
  \BibitemOpen
  \bibfield{author}{%
  \bibinfo {author} {\bibfnamefont{R.}~\bibnamefont{Gregory}}, \bibinfo
  {author} {\bibfnamefont{D.}~\bibnamefont{Kubiznak}},\ and\ \bibinfo {author}
  {\bibfnamefont{D.}~\bibnamefont{Wills}}}%
   (\bibinfo {year} {2013}),\
  \Eprint{http://arxiv.org/abs/1303.0519}{arXiv:1303.0519 [gr-qc]}%
  \bibAnnoteFile{NoStop}{Gregory:2013xca}%
\bibitem{Berti:2013gfa}%
  \BibitemOpen
  \bibfield{author}{%
  \bibinfo {author} {\bibfnamefont{E.}~\bibnamefont{Berti}}, \bibinfo {author}
  {\bibfnamefont{V.}~\bibnamefont{Cardoso}}, \bibinfo {author}
  {\bibfnamefont{L.}~\bibnamefont{Gualtieri}}, \bibinfo {author}
  {\bibfnamefont{M.}~\bibnamefont{Horbatsch}},\ and\ \bibinfo {author}
  {\bibfnamefont{U.}~\bibnamefont{Sperhake}}}%
   (\bibinfo {year} {2013}),\
  \Eprint{http://arxiv.org/abs/1304.2836}{arXiv:1304.2836 [gr-qc]}%
  \bibAnnoteFile{NoStop}{Berti:2013gfa}%
\bibitem{Blanchet:1995fg}%
  \BibitemOpen
  \bibfield{author}{%
  \bibinfo {author} {\bibfnamefont{L.}~\bibnamefont{Blanchet}}, \bibinfo
  {author} {\bibfnamefont{T.}~\bibnamefont{Damour}},\ and\ \bibinfo {author}
  {\bibfnamefont{B.~R.}\ \bibnamefont{Iyer}},\ }%
  \bibfield{journal}{%
  \Doi{10.1103/PhysRevD.51.5360, 10.1103/PhysRevD.54.1860}{\bibinfo {journal}
  {Phys.Rev.}}\ }%
  \textbf{\bibinfo {volume} {D51}},\ \bibinfo {pages} {5360} (\bibinfo {year}
  {1995}),\ \Eprint{http://arxiv.org/abs/gr-qc/9501029}{arXiv:gr-qc/9501029
  [gr-qc]}%
  \bibAnnoteFile{NoStop}{Blanchet:1995fg}%
\bibitem{Kainulainen:2006wz}%
  \BibitemOpen
  \bibfield{author}{%
  \bibinfo {author} {\bibfnamefont{K.}~\bibnamefont{Kainulainen}}, \bibinfo
  {author} {\bibfnamefont{V.}~\bibnamefont{Reijonen}},\ and\ \bibinfo {author}
  {\bibfnamefont{D.}~\bibnamefont{Sunhede}},\ }%
  \bibfield{journal}{%
  \Doi{10.1103/PhysRevD.76.043503}{\bibinfo {journal} {Phys.Rev.}}\ }%
  \textbf{\bibinfo {volume} {D76}},\ \bibinfo {pages} {043503} (\bibinfo {year}
  {2007}),\ \Eprint{http://arxiv.org/abs/gr-qc/0611132}{arXiv:gr-qc/0611132
  [gr-qc]}%
  \bibAnnoteFile{NoStop}{Kainulainen:2006wz}%
\bibitem{Barausse:2007pn}%
  \BibitemOpen
  \bibfield{author}{%
  \bibinfo {author} {\bibfnamefont{E.}~\bibnamefont{Barausse}}, \bibinfo
  {author} {\bibfnamefont{T.~P.}\ \bibnamefont{Sotiriou}},\ and\ \bibinfo
  {author} {\bibfnamefont{J.~C.}\ \bibnamefont{Miller}},\ }%
  \bibfield{journal}{%
  \Doi{10.1088/0264-9381/25/6/062001}{\bibinfo {journal} {Class.Quant.Grav.}}\
  }%
  \textbf{\bibinfo {volume} {25}},\ \bibinfo {pages} {062001} (\bibinfo {year}
  {2008}),\ \Eprint{http://arxiv.org/abs/gr-qc/0703132}{arXiv:gr-qc/0703132
  [GR-QC]}%
  \bibAnnoteFile{NoStop}{Barausse:2007pn}%
\bibitem{Olmo:2008pv}%
  \BibitemOpen
  \bibfield{author}{%
  \bibinfo {author} {\bibfnamefont{G.~J.}\ \bibnamefont{Olmo}},\ }%
  \bibfield{journal}{%
  \Doi{10.1103/PhysRevD.78.104026}{\bibinfo {journal} {Phys.Rev.}}\ }%
  \textbf{\bibinfo {volume} {D78}},\ \bibinfo {pages} {104026} (\bibinfo {year}
  {2008}),\ \Eprint{http://arxiv.org/abs/0810.3593}{arXiv:0810.3593 [gr-qc]}%
  \bibAnnoteFile{NoStop}{Olmo:2008pv}%
\bibitem{Olmo:2011np}%
  \BibitemOpen
  \bibfield{author}{%
  \bibinfo {author} {\bibfnamefont{G.~J.}\ \bibnamefont{Olmo}}\ and\ \bibinfo
  {author} {\bibfnamefont{D.}~\bibnamefont{Rubiera-Garcia}},\ }%
  \bibfield{journal}{%
  \Doi{10.1140/epjc/s10052-012-2098-7}{\bibinfo {journal} {Eur.Phys.J.}}\ }%
  \textbf{\bibinfo {volume} {C72}},\ \bibinfo {pages} {2098} (\bibinfo {year}
  {2012}),\ \Eprint{http://arxiv.org/abs/1112.0475}{arXiv:1112.0475 [gr-qc]}%
  \bibAnnoteFile{NoStop}{Olmo:2011np}%
\bibitem{Olmo:2012nx}%
  \BibitemOpen
  \bibfield{author}{%
  \bibinfo {author} {\bibfnamefont{G.~J.}\ \bibnamefont{Olmo}}\ and\ \bibinfo
  {author} {\bibfnamefont{D.}~\bibnamefont{Rubiera-Garcia}},\ }%
  \bibfield{journal}{%
  \Doi{10.1103/PhysRevD.86.044014}{\bibinfo {journal} {Phys.Rev.}}\ }%
  \textbf{\bibinfo {volume} {D86}},\ \bibinfo {pages} {044014} (\bibinfo {year}
  {2012}),\ \Eprint{http://arxiv.org/abs/1207.6004}{arXiv:1207.6004 [gr-qc]}%
  \bibAnnoteFile{NoStop}{Olmo:2012nx}%
\bibitem{Olmo:2012er}%
  \BibitemOpen
  \bibfield{author}{%
  \bibinfo {author} {\bibfnamefont{G.~J.}\ \bibnamefont{Olmo}}, \bibinfo
  {author} {\bibfnamefont{H.}~\bibnamefont{Sanchis-Alepuz}},\ and\ \bibinfo
  {author} {\bibfnamefont{S.}~\bibnamefont{Tripathi}},\ }%
  \bibfield{journal}{%
  \Doi{10.1103/PhysRevD.86.104039}{\bibinfo {journal} {Phys.Rev.}}\ }%
  \textbf{\bibinfo {volume} {D86}},\ \bibinfo {pages} {104039} (\bibinfo {year}
  {2012}),\ \Eprint{http://arxiv.org/abs/1211.0692}{arXiv:1211.0692 [gr-qc]}%
  \bibAnnoteFile{NoStop}{Olmo:2012er}%
\bibitem{Li:2008bma}%
  \BibitemOpen
  \bibfield{author}{%
  \bibinfo {author} {\bibfnamefont{B.}~\bibnamefont{Li}}, \bibinfo {author}
  {\bibfnamefont{D.~F.}\ \bibnamefont{Mota}},\ and\ \bibinfo {author}
  {\bibfnamefont{D.~J.}\ \bibnamefont{Shaw}},\ }%
  \bibfield{journal}{%
  \Doi{10.1088/0264-9381/26/5/055018}{\bibinfo {journal} {Class.Quant.Grav.}}\
  }%
  \textbf{\bibinfo {volume} {26}},\ \bibinfo {pages} {055018} (\bibinfo {year}
  {2009}),\ \Eprint{http://arxiv.org/abs/0801.0603}{arXiv:0801.0603 [gr-qc]}%
  \bibAnnoteFile{NoStop}{Li:2008bma}%
\bibitem{Li:2008fa}%
  \BibitemOpen
  \bibfield{author}{%
  \bibinfo {author} {\bibfnamefont{B.}~\bibnamefont{Li}}, \bibinfo {author}
  {\bibfnamefont{D.~F.}\ \bibnamefont{Mota}},\ and\ \bibinfo {author}
  {\bibfnamefont{D.~J.}\ \bibnamefont{Shaw}},\ }%
  \bibfield{journal}{%
  \Doi{10.1103/PhysRevD.78.064018}{\bibinfo {journal} {Phys.Rev.}}\ }%
  \textbf{\bibinfo {volume} {D78}},\ \bibinfo {pages} {064018} (\bibinfo {year}
  {2008}),\ \Eprint{http://arxiv.org/abs/0805.3428}{arXiv:0805.3428 [gr-qc]}%
  \bibAnnoteFile{NoStop}{Li:2008fa}%
\bibitem{Koivisto:2005yc}%
  \BibitemOpen
  \bibfield{author}{%
  \bibinfo {author} {\bibfnamefont{T.}~\bibnamefont{Koivisto}}\ and\ \bibinfo
  {author} {\bibfnamefont{H.}~\bibnamefont{Kurki-Suonio}},\ }%
  \bibfield{journal}{%
  \Doi{10.1088/0264-9381/23/7/009}{\bibinfo {journal} {Class.Quant.Grav.}}\ }%
  \textbf{\bibinfo {volume} {23}},\ \bibinfo {pages} {2355} (\bibinfo {year}
  {2006}),\
  \Eprint{http://arxiv.org/abs/astro-ph/0509422}{arXiv:astro-ph/0509422
  [astro-ph]}%
  \bibAnnoteFile{NoStop}{Koivisto:2005yc}%
\bibitem{Koivisto:2006ie}%
  \BibitemOpen
  \bibfield{author}{%
  \bibinfo {author} {\bibfnamefont{T.}~\bibnamefont{Koivisto}},\ }%
  \bibfield{journal}{%
  \Doi{10.1103/PhysRevD.73.083517}{\bibinfo {journal} {Phys.Rev.}}\ }%
  \textbf{\bibinfo {volume} {D73}},\ \bibinfo {pages} {083517} (\bibinfo {year}
  {2006}),\
  \Eprint{http://arxiv.org/abs/astro-ph/0602031}{arXiv:astro-ph/0602031
  [astro-ph]}%
  \bibAnnoteFile{NoStop}{Koivisto:2006ie}%
\bibitem{Koivisto:2007sq}%
  \BibitemOpen
  \bibfield{author}{%
  \bibinfo {author} {\bibfnamefont{T.}~\bibnamefont{Koivisto}},\ }%
  \bibfield{journal}{%
  \Doi{10.1103/PhysRevD.76.043527}{\bibinfo {journal} {Phys.Rev.}}\ }%
  \textbf{\bibinfo {volume} {D76}},\ \bibinfo {pages} {043527} (\bibinfo {year}
  {2007}),\ \Eprint{http://arxiv.org/abs/0706.0974}{arXiv:0706.0974
  [astro-ph]}%
  \bibAnnoteFile{NoStop}{Koivisto:2007sq}%
\bibitem{Amendola:2010bk}%
  \BibitemOpen
  \bibfield{author}{%
  \bibinfo {author} {\bibfnamefont{L.}~\bibnamefont{Amendola}}, \bibinfo
  {author} {\bibfnamefont{K.}~\bibnamefont{Enqvist}},\ and\ \bibinfo {author}
  {\bibfnamefont{T.}~\bibnamefont{Koivisto}},\ }%
  \bibfield{journal}{%
  \Doi{10.1103/PhysRevD.83.044016}{\bibinfo {journal} {Phys.Rev.}}\ }%
  \textbf{\bibinfo {volume} {D83}},\ \bibinfo {pages} {044016} (\bibinfo {year}
  {2011}),\ \Eprint{http://arxiv.org/abs/1010.4776}{arXiv:1010.4776 [gr-qc]}%
  \bibAnnoteFile{NoStop}{Amendola:2010bk}%
\bibitem{Koivisto:2011vq}%
  \BibitemOpen
  \bibfield{author}{%
  \bibinfo {author} {\bibfnamefont{T.~S.}\ \bibnamefont{Koivisto}},\ }%
  \bibfield{journal}{%
  \Doi{10.1103/PhysRevD.83.101501}{\bibinfo {journal} {Phys.Rev.}}\ }%
  \textbf{\bibinfo {volume} {D83}},\ \bibinfo {pages} {101501} (\bibinfo {year}
  {2011}),\ \Eprint{http://arxiv.org/abs/1103.2743}{arXiv:1103.2743 [gr-qc]}%
  \bibAnnoteFile{NoStop}{Koivisto:2011vq}%
\bibitem{Harko:2011nh}%
  \BibitemOpen
  \bibfield{author}{%
  \bibinfo {author} {\bibfnamefont{T.}~\bibnamefont{Harko}}, \bibinfo {author}
  {\bibfnamefont{T.~S.}\ \bibnamefont{Koivisto}}, \bibinfo {author}
  {\bibfnamefont{F.~S.}\ \bibnamefont{Lobo}},\ and\ \bibinfo {author}
  {\bibfnamefont{G.~J.}\ \bibnamefont{Olmo}},\ }%
  \bibfield{journal}{%
  \Doi{10.1103/PhysRevD.85.084016}{\bibinfo {journal} {Phys.Rev.}}\ }%
  \textbf{\bibinfo {volume} {D85}},\ \bibinfo {pages} {084016} (\bibinfo {year}
  {2012}),\ \Eprint{http://arxiv.org/abs/1110.1049}{arXiv:1110.1049 [gr-qc]}%
  \bibAnnoteFile{NoStop}{Harko:2011nh}%
\bibitem{Sandstad:2013oja}%
  \BibitemOpen
  \bibfield{author}{%
  \bibinfo {author} {\bibfnamefont{M.}~\bibnamefont{Sandstad}}, \bibinfo
  {author} {\bibfnamefont{T.~S.}\ \bibnamefont{Koivisto}},\ and\ \bibinfo
  {author} {\bibfnamefont{D.~F.}\ \bibnamefont{Mota}},\ }%
  \bibfield{journal}{%
  \bibinfo {journal} {Class. Quantum Grav. 30,}\ }%
  \textbf{\bibinfo {volume} {155005}} (\bibinfo {year} {2013}),\ \doi{\bibinfo
  {doi} {10.1088/0264-9381/30/15/155005}},\
  \Eprint{http://arxiv.org/abs/1305.0695}{arXiv:1305.0695 [gr-qc]}%
  \bibAnnoteFile{NoStop}{Sandstad:2013oja}%
\bibitem{Koivisto:2011tp}%
  \BibitemOpen
  \bibfield{author}{%
  \bibinfo {author} {\bibfnamefont{T.~S.}\ \bibnamefont{Koivisto}},\ }%
  \bibfield{journal}{%
  \Doi{10.1103/PhysRevD.84.121502}{\bibinfo {journal} {Phys.Rev.}}\ }%
  \textbf{\bibinfo {volume} {D84}},\ \bibinfo {pages} {121502} (\bibinfo {year}
  {2011}),\ \Eprint{http://arxiv.org/abs/1109.4585}{arXiv:1109.4585 [gr-qc]}%
  \bibAnnoteFile{NoStop}{Koivisto:2011tp}%
\bibitem{wiseman}%
  \BibitemOpen
  \bibfield{author}{%
  \bibinfo {author} {\bibfnamefont{C.~M.}\ \bibnamefont{Will}}\ and\ \bibinfo
  {author} {\bibfnamefont{A.~G.}\ \bibnamefont{Wiseman}},\ }%
  \bibfield{journal}{%
  \Doi{10.1103/PhysRevD.54.4813}{\bibinfo {journal} {Phys. Rev. D}}\ }%
  \textbf{\bibinfo {volume} {54}},\ \bibinfo {pages} {4813} (\bibinfo {year}
  {1996}),\ \Eprint{http://arxiv.org/abs/gr-qc/9608012}{arXiv:gr-qc/9608012},\
  \url{http://link.aps.org/doi/10.1103/PhysRevD.54.4813}%
  \bibAnnoteFile{NoStop}{wiseman}%
\bibitem{pati1}%
  \BibitemOpen
  \bibfield{author}{%
  \bibinfo {author} {\bibfnamefont{M.~E.}\ \bibnamefont{Pati}}\ and\ \bibinfo
  {author} {\bibfnamefont{C.~M.}\ \bibnamefont{Will}},\ }%
  \bibfield{journal}{%
  \Doi{10.1103/PhysRevD.62.124015}{\bibinfo {journal} {Phys. Rev. D}}\ }%
  \textbf{\bibinfo {volume} {62}},\ \bibinfo {pages} {124015} (\bibinfo {year}
  {2000}),\ \Eprint{http://arxiv.org/abs/gr-qc/0007087}{arXiv:gr-qc/0007087},\
  \url{http://link.aps.org/doi/10.1103/PhysRevD.62.124015}%
  \bibAnnoteFile{NoStop}{pati1}%
\bibitem{Koivisto:2005yk}%
  \BibitemOpen
  \bibfield{author}{%
  \bibinfo {author} {\bibfnamefont{T.}~\bibnamefont{Koivisto}},\ }%
  \bibfield{journal}{%
  \Doi{10.1088/0264-9381/23/12/N01}{\bibinfo {journal} {Class.Quant.Grav.}}\ }%
  \textbf{\bibinfo {volume} {23}},\ \bibinfo {pages} {4289} (\bibinfo {year}
  {2006}),\ \Eprint{http://arxiv.org/abs/gr-qc/0505128}{arXiv:gr-qc/0505128
  [gr-qc]}%
  \bibAnnoteFile{NoStop}{Koivisto:2005yk}%
\bibitem{gravitation}%
  \BibitemOpen
  \bibfield{author}{%
  \bibinfo {author} {\bibfnamefont{C.~W.}\ \bibnamefont{{Misner}}}, \bibinfo
  {author} {\bibfnamefont{K.~S.}\ \bibnamefont{{Thorne}}},\ and\ \bibinfo
  {author} {\bibfnamefont{J.~A.}\ \bibnamefont{{Wheeler}}},\ }%
  \emph{\bibinfo {title} {{Gravitation}}}\ (\bibinfo {publisher} {W. H. Freeman
  and Co.},\ \bibinfo {address} {New York},\ \bibinfo {year} {1973})%
  \bibAnnoteFile{NoStop}{gravitation}%
\bibitem{pati2}%
  \BibitemOpen
  \bibfield{author}{%
  \bibinfo {author} {\bibfnamefont{M.~E.}\ \bibnamefont{Pati}}\ and\ \bibinfo
  {author} {\bibfnamefont{C.~M.}\ \bibnamefont{Will}},\ }%
  \bibfield{journal}{%
  \Doi{10.1103/PhysRevD.65.104008}{\bibinfo {journal} {Phys. Rev. D}}\ }%
  \textbf{\bibinfo {volume} {65}},\ \bibinfo {pages} {104008} (\bibinfo {year}
  {2002}),\ \Eprint{http://arxiv.org/abs/gr-qc/9910057}{arXiv:gr-qc/9910057},\
  \url{http://link.aps.org/doi/10.1103/PhysRevD.65.104008}%
  \bibAnnoteFile{NoStop}{pati2}%
\bibitem{Capozziello:2012eu}%
  \BibitemOpen
  \bibfield{author}{%
  \bibinfo {author} {\bibfnamefont{S.}~\bibnamefont{Capozziello}}, \bibinfo
  {author} {\bibfnamefont{M.}~\bibnamefont{De~Laurentis}}, \bibinfo {author}
  {\bibfnamefont{L.}~\bibnamefont{Fatibene}},\ and\ \bibinfo {author}
  {\bibfnamefont{M.}~\bibnamefont{Francaviglia}},\ }%
  \bibfield{journal}{%
  \Doi{10.1142/S0219887812500727}{\bibinfo {journal}
  {Int.J.Geom.Meth.Mod.Phys.}}\ }%
  \textbf{\bibinfo {volume} {9}},\ \bibinfo {pages} {1250072} (\bibinfo {year}
  {2012}),\ \Eprint{http://arxiv.org/abs/1202.5699}{arXiv:1202.5699 [gr-qc]}%
  \bibAnnoteFile{NoStop}{Capozziello:2012eu}%
\bibitem{Olmo:2005zr}%
  \BibitemOpen
  \bibfield{author}{%
  \bibinfo {author} {\bibfnamefont{G.~J.}\ \bibnamefont{Olmo}},\ }%
  \bibfield{journal}{%
  \Doi{10.1103/PhysRevLett.95.261102}{\bibinfo {journal} {Phys.Rev.Lett.}}\ }%
  \textbf{\bibinfo {volume} {95}},\ \bibinfo {pages} {261102} (\bibinfo {year}
  {2005}),\ \Eprint{http://arxiv.org/abs/gr-qc/0505101}{arXiv:gr-qc/0505101
  [gr-qc]}%
  \bibAnnoteFile{NoStop}{Olmo:2005zr}%
\bibitem{Flanagan:2003rb}%
  \BibitemOpen
  \bibfield{author}{%
  \bibinfo {author} {\bibfnamefont{E.~E.}\ \bibnamefont{Flanagan}},\ }%
  \bibfield{journal}{%
  \Doi{10.1103/PhysRevLett.92.071101}{\bibinfo {journal} {Phys.Rev.Lett.}}\ }%
  \textbf{\bibinfo {volume} {92}},\ \bibinfo {pages} {071101} (\bibinfo {year}
  {2004}),\
  \Eprint{http://arxiv.org/abs/astro-ph/0308111}{arXiv:astro-ph/0308111
  [astro-ph]}%
  \bibAnnoteFile{NoStop}{Flanagan:2003rb}%
\bibitem{Vollick:2003ic}%
  \BibitemOpen
  \bibfield{author}{%
  \bibinfo {author} {\bibfnamefont{D.~N.}\ \bibnamefont{Vollick}},\ }%
  \bibfield{journal}{%
  \Doi{10.1088/0264-9381/21/15/N01}{\bibinfo {journal} {Class.Quant.Grav.}}\ }%
  \textbf{\bibinfo {volume} {21}},\ \bibinfo {pages} {3813} (\bibinfo {year}
  {2004}),\ \Eprint{http://arxiv.org/abs/gr-qc/0312041}{arXiv:gr-qc/0312041
  [gr-qc]}%
  \bibAnnoteFile{NoStop}{Vollick:2003ic}%
\bibitem{Olmo:2006zu}%
  \BibitemOpen
  \bibfield{author}{%
  \bibinfo {author} {\bibfnamefont{G.~J.}\ \bibnamefont{Olmo}},\ }%
  \bibfield{journal}{%
  \Doi{10.1103/PhysRevLett.98.061101}{\bibinfo {journal} {Phys.Rev.Lett.}}\ }%
  \textbf{\bibinfo {volume} {98}},\ \bibinfo {pages} {061101} (\bibinfo {year}
  {2007}),\ \Eprint{http://arxiv.org/abs/gr-qc/0612002}{arXiv:gr-qc/0612002
  [gr-qc]}%
  \bibAnnoteFile{NoStop}{Olmo:2006zu}%
\end{thebibliography}%

\end{document}